# Gradient Electronic Landscapes in van der Waals Heterostructures

*Nolan Lassaline,[*,†] Camilla H. Sørensen,[†] Giulia Meucci,[§] Sander J. Linde,[†] Kian Latifi Yaghin,[†] Tuan K. Chau,[†] Damon J. Carrad,[§] Peter Bøggild,[†] Thomas S. Jespersen,[§] Timothy J. Booth[†]*

[†]Department of Physics, Technical University of Denmark, 2800 Kongens Lyngby, Denmark
[§]Department of Energy Conversion and Storage, Technical University of Denmark, 2800 Kongens Lyngby, Denmark

ABSTRACT. Two-dimensional (2D) materials such as graphene and hexagonal boron nitride (hBN) provide a versatile platform for quantum electronics. Experiments generally require encapsulating graphene within hBN flakes, forming a protective van der Waals (vdW) heterostructure that preserves delicate properties of the embedded crystal. To produce functional devices, heterostructures are typically shaped by electron beam lithography and etching, which has driven progress in 2D materials research. However, patterns are primarily restricted to in-plane geometries such as boxes, holes, and stripes, limiting opportunities for advanced architectures. Here, we use thermal scanning-probe lithography (tSPL) to produce smooth topographic landscapes in vdW heterostructures, controlling the thickness degree of freedom with nanometer precision. We electrically gate a sinusoidal topography to impose an electric-field gradient on the graphene layer to spatially modulate charge-carrier doping. We observe signatures of the landscape in transport measurements—resistance-peak spreading and commensurability oscillations—establishing tSPL for tailoring high-quality quantum electronics.

Two-dimensional (2D) materials are atomically thin sheets with reduced dimensionality and quantum confinement effects[1]. 2D crystals such as graphene[2], hexagonal boron nitride (hBN)[3], and transition metal dichalcogenides (TMDs)[4] exhibit remarkable properties suitable for a broad range of applications[5]. The high surface-area-to-volume ratio of 2D materials, combined with quantum confinement of charge carriers, enables strong interactions with external fields that can be tailored by modifying the local environment. The tunability and emerging properties of 2D materials have established them at the forefront in applications such as quantum electronics[6–9], nanophotonics[10–13], and optoelectronic technologies[14–16].

To preserve and enhance the intrinsic properties of atomically thin materials, individual sheets can be stacked to form layered assemblies known as van der Waals (vdW) heterostructures[1,17]. Forming heterostructures with clean interfaces is critical, as the encapsulation of delicate crystals protects their pristine properties from degradation and unwanted environmental influence. Furthermore, vdW stacking enables the combination of hybrid material classes at atomically precise interfaces, offering a versatile and robust platform for engineering multifunctional devices[18–21]. Beyond stacking, vdW heterostructures can be further modified using lithographic techniques to create patterned devices with functionalities exceeding those intrinsic to the materials[22–24].

The conventional strategy for patterning vdW heterostructures involves using electron beam lithography (EBL) to expose a polymer resist, followed by development and pattern transfer with an etching process[25]. This approach has been instrumental in the research and development of 2D materials and their devices, however, EBL imposes a strong limitation on pattern geometries. Because the lithographic process is defined strictly in-plane, it produces "binary" profiles where the resist is fully exposed or not exposed at all, limiting available geometries to two-dimensional arrays. Grayscale EBL has been explored to control nanopatterning on the vertical axis, yet it remains a difficult challenge to control surface topography with nanometer precision[26]. Researchers are exploring approaches to increase control over the surface profile; however, structures remain primarily limited to binary shapes[27–29]. Further research would benefit from more sophisticated device architectures[30–32], and there is a growing need for methods that can produce freeform contours to generate out-of-plane electric-field gradients[33–36].

In this study, we use thermal scanning-probe lithography (tSPL)[37–40] and reactive ion etching (RIE) to produce grayscale surface topographies[41–44] in hBN flakes[45,46], allowing us to create gated structures that impose spatially varying out-of-plane electric fields, $E_{\perp}(r)$, on the encapsulated graphene layer[47]. We call this concept gradient electronic landscapes since $E_{\perp}(r)$ has a non-zero gradient, $\nabla E_{\perp}(r) \neq 0$, generated by the patterned surface. In a standard flat device, electrical gates are used to tune material properties to a single value throughout the device or under patterned gate regions with constant thickness. In contrast, our device allows material properties to be continuously modulated along the in-plane spatial axes, enabling the realization of mathematically precise doping landscapes in an embedded 2D layer without patterning or damaging the crystal. The gated topographic landscape spatially modulates carrier density, offering enhanced freedom to tailor charge carriers in a vdW heterostructure. We implement this concept to tailor ballistic transport for dirac fermions in graphene, demonstrating the use of tSPL for modifying electronic behavior in quantum systems beyond conventional limits.

tSPL works by scanning an electrically heated silicon tip (Figure 1a) over a polymer layer (polyphthalaldehyde, PPA), and by controlling the temperature and applied downward force of the tip, PPA is locally sublimated to sculpt the surface through material removal[48]. The probe is raster-scanned to pattern each pixel for a given structure, such that a smooth topographic landscape defined by a mathematical function is produced (Figure 1b). The in-plane resolution is on the order of 10 nm[49] and the vertical resolution is on the order of 1 nm[50], leading to continuous surface profiles with low residual error[40–46]. tSPL typically operates in ambient conditions, where the pixel patterning rate (40 kHz) ensures fast operation on the order of a few minutes for patterns with an area of up to 50×50 $\mu m^2$. Furthermore, tSPL can read the surface topography of the PPA layer on 2D materials before, during, and after patterning, offering a powerful environment for combined lithography and characterization with rapid turnaround time and nanometer-precise reading and writing capabilities.

In this work, we introduce a new device concept based on using tSPL to pattern the top flake of a hBN–graphene–hBN heterostructure, followed by thermal evaporation of a gold layer on top to produce a topographically structured top gate (Figure 1c). This approach takes advantage of the unique ability of tSPL to pattern grayscale topographies in a resist layer with nanoscale resolution, and by transferring these patterns with reactive ion etching (RIE), the surface profile is accurately reproduced in hBN with nanometer precision[51] (Figure 1d–g). This approach offers a simple and powerful route to tailor electronic landscapes in vdW heterostructures.

2D crystals are prepared by mechanical exfoliation of graphene and hBN into monolayers and multilayer flakes, respectively, and placed on silicon chips with ~285 nm $SiO_2$. Monolayer graphene and hBN flakes of ~15–70 nm thickness are identified using optical microscopy. The hBN thickness is chosen to be minimal, while still isolating the graphene from the underlying $SiO_2$ surface (bottom hBN, ~15 nm thickness), and while leaving enough room for a topographic landscape in the top hBN flake (~65 nm thickness). The crystals are then stacked to place the hBN–graphene–hBN heterostructure on an electrical chip, with thin hBN on the bottom and thick hBN on top (Supporting Information).

The sample is prepared for tSPL by spin-coating a layer of PPA with ~75 nm thickness. The chip is placed on the tSPL stage, and the position of the heterostructure is located using a built-in optical microscope. Calibration patterns are performed to optimize feedback and patterning conditions. The surface topography of spin-coated PPA on the heterostructure is measured to locate flat areas that are free of defects such as cracks, bubbles, or other non-uniformities. The probe also produces a small pattern with maximum depth to measure the local PPA thickness on the heterostructure, which is critical for accurately calibrating the etching process. The region for lithography is identified and the tool is ready for patterning. The desired topographic landscape is converted from a mathematical function to a bitmap with 20×20 $nm^2$ pixel size and 8-bit depth precision—in this work a sine wave with a period of ~300 nm and a peak-to-valley depth modulation of ~15 nm. The input bitmap is patterned as a topographic landscape in the PPA layer on top of the heterostructure, where the surface profile is measured during writing to adjust feedback parameters and record topographic data.

The PPA structure is transferred to the top hBN flake using RIE with $SF_6$ gas. It is critical to know the etch rate of the system and the PPA thickness from tSPL measurements, as the etching step should be accurately timed such that the entire depth range of the topographic landscape is transferred, but without etching through the flake[51]. After etching the sample is cleaned in a multi-step procedure including thermal, chemical, and physical processes. Details of the etching and cleaning procedure are provided in the Supporting Information.

The hBN topography is measured by atomic force microscopy (AFM) to quantify pattern fidelity and errors introduced by the fabrication process. Tapping-mode AFM is used to extract the height profile of the landscape and the location of the pattern on the stack. We use the *FunFit* software program[51] to fit a sinusoidal function to the measured data, revealing a wavelength of 296 nm and a modulation depth of 12 nm in hBN (Figure 1). The pattern fidelity is quantified through root-mean-square error (RMSE) analysis, where the error of the hBN surface pattern is 1.4 nm (1.3 nm in PPA before etch transfer). When compared to the interlayer spacing of hBN crystals (~0.33 nm), the error is comparable to just a few atomic distances, highlighting the precision of this approach for defining structured electrical gates.

The lateral region defining the top-gate boundary is produced by EBL in a ~200-nm-thick poly-methyl methacrylate (PMMA) layer, spin-coated on the patterned stack. The exposed pattern is developed, such that an opening in the polymer layer of ~3×5 $\mu m^2$ is produced on the sinusoidal surface, which is routed to a contact electrode on the perimeter of the chip. A Cr/Au layer of ~5 nm/~50 nm is thermally evaporated and removed using a liftoff process, such that only the top-gate region and connection electrode are metallized. In addition to defining the top gate, this step protects the heterostructure underneath the metal layer during subsequent process steps. The lateral device boundary is shaped by combining EBL with sequential RIE using $SF_6$ gas to etch through hBN and $O_2$ gas to etch through graphene. This step shapes the device into a Hall bar geometry and electrically isolates the conductive graphene plane from the sample surroundings. A final EBL step is performed to create electrical edge contacts to the graphene layer, which are routed to contact pads at the chip perimeter, followed by wire-bonding the sample to a carrier used for cryostat measurements (Supporting Information).

The heterostructure is now shaped into a Hall bar geometry for electrical measurements[2,22], where the new feature presented in this work is a top gate with a smooth topographic landscape. Current moves from source to drain ($I_{\mathrm{sd}}$) along the Hall bar, and the longitudinal ($V_{xx}$) and transverse ($V_{xy}$) voltages are measured in a four-point configuration in a dilution refrigerator with 15 mK base temperature[52], where Figure 2a shows a schematic of the measurement setup. The device is globally doped by applying a back-gate bias ($V_{\mathrm{bg}}$) between silicon and graphene, and local doping is controlled by applying a top-gate bias ($V_{\mathrm{tg}}$) between graphene and gold. Figure 2b shows an image of fabricated devices, where the device used in this study is indicated. Figure 2c shows a cross section of the device stack, illustrating how the patterned top gate and global back gate are biased to spatially tailor the charge carrier distribution. Figure 2d shows a collection of properties that are modulated along a spatial axis of the device ($x$), emerging from the sinusoidal top gate and global back gate. It is interesting to note that these properties typically take on a single value throughout a standard "flat" device, while here they become spatially distributed, such that the energy axis of the system unfolds onto a spatial axis. This results in a modulated electric potential, doping profile, resistance, and cyclotron radius in an applied magnetic field, which have observable effects that show up in the following electrical measurements.

Device operation depends on the applied gate voltages, which controls the carrier density, polarity, modulation strength, and cyclotron radius, and thereby the behaviour of ballistic charge carriers in the periodic electric field. Figure 3a shows the average carrier density in the device plotted as a function of $V_{\mathrm{bg}}$ and $V_{\mathrm{tg}}$, where the average is displayed due to the spatial modulation of the local carrier density along the $x$-coordinate, induced by the top gate (Figure 3a insets, Supporting Information). Charge neutrality ($n_{\mathrm{av}} = 0$) in the device follows a linear relationship on this map due to the combined influence of the top and back gates. The region to the right of this line represents *n*-doping (electrons), and the region to the left of the line represents *p*-doping (holes), forming 4 quadrants in the plot defined by unipolar regimes with one type of charge carrier (*nn'* or *pp'*, where *n'* and *p'* indicate a single carrier type with spatially modulated magnitude from $n$ to *n'* or $p$ to *p'*) or bipolar regimes with mixed charge carriers (*np* or *pn*). The insets show the carrier density as a function of space along the *x*-coordinate, normalized to the maximum and minimum, for a pair of gate voltages indicated by the yellow dots, which are exactly on the charge neutrality line for the bipolar quadrants as $n_{\mathrm{av}} = 0$ is in the middle of the sinusoid. The induced doping profile determines the operation regime of the device and the properties that arise from the spatial distribution of charge carriers in the system.

The longitudinal resistance $R_{xx}$ is extracted from measuring $V_{xx}$ under an AC current bias of $I_{\mathrm{sd}} = 200$ nA at $T \approx 15$ mK, as a function of $V_{\mathrm{bg}}$ and $V_{\mathrm{tg}}$, allowing us to experimentally map the device properties as the global and local doping are tuned by the back gate and top gate voltages. Figure 3b shows the measured $R_{xx}(V_{\mathrm{bg}}, V_{\mathrm{tg}})$, where the peak origin is located at $V'_{\mathrm{bg}} = V'_{\mathrm{tg}} = 0$, where $V'_{\mathrm{bg}} = V_{\mathrm{bg}} - V_{0,\mathrm{bg}}$ and $V'_{\mathrm{tg}} = V_{\mathrm{tg}} - V_{0,\mathrm{tg}}$, due to offsets $V_{0,\mathrm{bg}} = 2.4$ V and $V_{0,\mathrm{tg}} = -1$ V possibly caused by charge-impurity doping. A line cut of this plot at $V'_{\mathrm{tg}} = 0$ reveals a peak in $R_{xx}$ at $V'_{\mathrm{bg}} = 0$ due to charge neutrality, as expected for a heterostructure with only a back gate (Figure S9). Analysis of the resistance peak at $V'_{\mathrm{tg}} = 0$ reveals field-effect mobilities for electrons and holes of $\mu_{\mathrm{e}} \approx 86{,}000\ \frac{\mathrm{cm}^2}{\mathrm{V}\cdot s}$ and $\mu_{\mathrm{h}} \approx 82{,}000\ \frac{\mathrm{cm}^2}{\mathrm{V}\cdot s}$, respectively (Supporting Information).

When the magnitude of the top gate voltage is increased, the measured $R_{xx}$ along $V'_{\mathrm{tg}} \neq 0$ reveals that the resistance peak no longer exists at a single point, but rather it spreads over a range of $V_{\mathrm{bg}}$ values, which we term here as resistance-peak spreading. The spread arises because charge neutrality—and therefore the resistance peak—exists no longer everywhere in the device for a single $V_{\mathrm{bg}}$, as in a flat device, but rather at a range of positions along $x$ for a range of $V_{\mathrm{bg}}$. This range is determined by the span of top-gate hBN thicknesses between minimum and maximum, which we confirm by overlaying calculated bounds for the charge neutrality peak in Figure 3b (red dashed lines, Supporting Information). Resistance-peak spreading is a key signature of the electric field gradient in our device, confirming the effect of a vertically patterned top gate.

A magnetic field with magnitude $B$ is now applied perpendicular to the graphene sheet, where Figure 4a shows the measured $R_{xx}$ as a function of $V_{\mathrm{bg}}$ and $B$, when the top gate is set to the charge neutrality point $V'_{\mathrm{tg}} = 0$. The data reveals a Landau fan originating from the charge neutrality point centered at $V_{\mathrm{bg}} = V_{0,\mathrm{bg}}$, as expected for gated monolayer graphene in a perpendicular magnetic field[22,53]. Figure 4b shows the measured $R_{xx}$ when the top gate is set to $V'_{\mathrm{tg}} = 9$ V, where two important differences are observed. First, the origin of the Landau fan is shifted to $V_{\mathrm{bg}} \approx -45$ V, which can be explained by the position of charge neutrality depending on the combined influence of both gates, as shown in Figure 3b. The second difference is the emergence of a set of bands oscillating between high and low resistance along the magnetic field axis for $B \leq 3$ T. When charge carriers in such a device are subjected to a periodic electric field imposed by the gate and an external perpendicular magnetic field, the measured longitudinal resistance exhibits a variation in magnitude known as commensurability oscillations (COs, also referred to as Weiss oscillations)[54–57], arising from the interplay between cyclotron orbits and a periodic electric field.

We note that a spurious resistance peak in Figure 4b located at $V_{\mathrm{bg}} \approx V_{0,\mathrm{bg}}$ is not an emerging feature compared to Figure 4a, it simply becomes more apparent as charge neutrality shifts away from $V_{\mathrm{bg}} = V_{0,\mathrm{bg}}$. Furthermore, this peak is weak compared to the main resistance peak, and its origin is likely due to resistance at contact edges, as the magnitude responded only to changing contact configuration, while the position remained fixed. Here, we adjusted the contact configuration to minimize the peak magnitude, and despite its persistence, the Landau fan and COs remain unaffected away from the peak.

Figure 4c shows a line cut of $R_{xx}$ along the $B$-axis for $V_{\mathrm{bg}} = 45$ V and $V'_{\mathrm{tg}} = 9$ V. Two sets of oscillations can be observed: Shubnikov–de Haas oscillations (SdHOs)[9] at high magnetic fields, and COs at low magnetic fields. The COs arise due to the interplay between the cyclotron radius and the imposed periodic electric field, where the commensurability condition is given by:

$$2R_{\mathrm{c}} = \left(\lambda - \frac{1}{4}\right) a$$

where $R_{\mathrm{c}}$ is the cyclotron radius, $a$ is the periodicity, and $\lambda$ is an integer[54,55]. The expected locations of the commensurability dips are indicated by the calculated yellow vertical lines (Supporting Information), showing excellent agreement with the measurement. We observe up to six dips in the data, indicating a clean heterostructure where ballistic transport is maintained over multiple periods of the landscape, comparing well with literature[22,55,56]. The inset shows a zoom-in of Figure 4b, where dips in $R_{xx}$ are marked with dashed yellow lines, highlighting the expected $B \propto \sqrt{n}$ dependence for COs[22], confirming their existence in the device (Supporting Information).

In this study we used tSPL to pattern smooth topographic surfaces in vdW heterostructures to spatially tailor the vertical electric field with a smooth gradient in an encapsulated graphene monolayer. We fabricated hBN–graphene–hBN heterostructures with a sinusoidal topography using a mixed approach[58] by combining tSPL, EBL, thermal evaporation, and RIE. The topography was measured by AFM and analyzed using the open-source *FunFit* software package, revealing nanometer-precise landscaping in hBN with a RMSE of 1.4 nm. When a top-gate voltage was applied, the topography imposed a sinusoidal electric field at the graphene layer. In contrast to standard devices with flat geometries, our device contains a gradient landscape that spatially modulates charge-carrier doping. We observe the electrostatic effect of the landscape in a measured dual-gate map, which shows resistance-peak spreading as a function of applied top-gate voltage. When subjected to a perpendicular magnetic field, electric current in the device exhibits commensurability oscillations in the measured longitudinal resistance, confirming the emergence of tailored transport induced by the imposed landscape. Our results demonstrate that precisely controlling the local thickness of vdW heterostructures is a powerful means for manipulating charge carriers in 2D materials, where more complex topographic landscapes can now be explored. The freedom offered by tSPL for spatially tailoring electronic properties unlocks interesting possibilities for ballistic electron optics in programmable *p–n* landscapes[59], moiré physics[60,61], quasiperiodic lattices[62], and gate-tunable quantum phases in space and time[30,63]. Beyond electronics, this approach can be exploited to tailor materials for optical[64], optoelectronic[65], and nanofluidic[66] applications.

ASSOCIATED CONTENT

**Supporting Information.** The Supporting Information is available free of charge on the ACS Publications website at https://pubs.acs.org.

AUTHOR INFORMATION

**Corresponding Author**

*Email: nlasso@dtu.dk

**ORCID**

Nolan Lassaline: 0000-0002-5854-3900

Camilla H. Sørensen: 0009-0005-8972-2185

Giulia Meucci: 0009-0007-4564-7835

Sander J. Linde: 0009-0004-0297-1836

Kian Latifi Yaghin: 0009-0008-2645-9406

Damon J. Carrad: 0000-0003-0372-8593

Peter Bøggild: 0000-0002-4342-0449

Thomas S. Jespersen: 0000-0002-7879-976X

Timothy J. Booth: 0000-0002-9784-989X

**Notes**

The authors declare no competing financial interest.

ACKNOWLEDGMENTS

N. L. acknowledges funding from the Swiss National Science Foundation (*Postdoc Mobility* P500PT_211105) and the Villum Foundation (*Villum Experiment* 50355). T. J. B. and P. B. acknowledge support from the Novo Nordisk Foundation (BIOMAG NNF21OC0066526). The authors thank D. H. Nguyen for assistance with flake preparation and thank R. Linde and I. Linde for stimulating discussions.

FIGURES

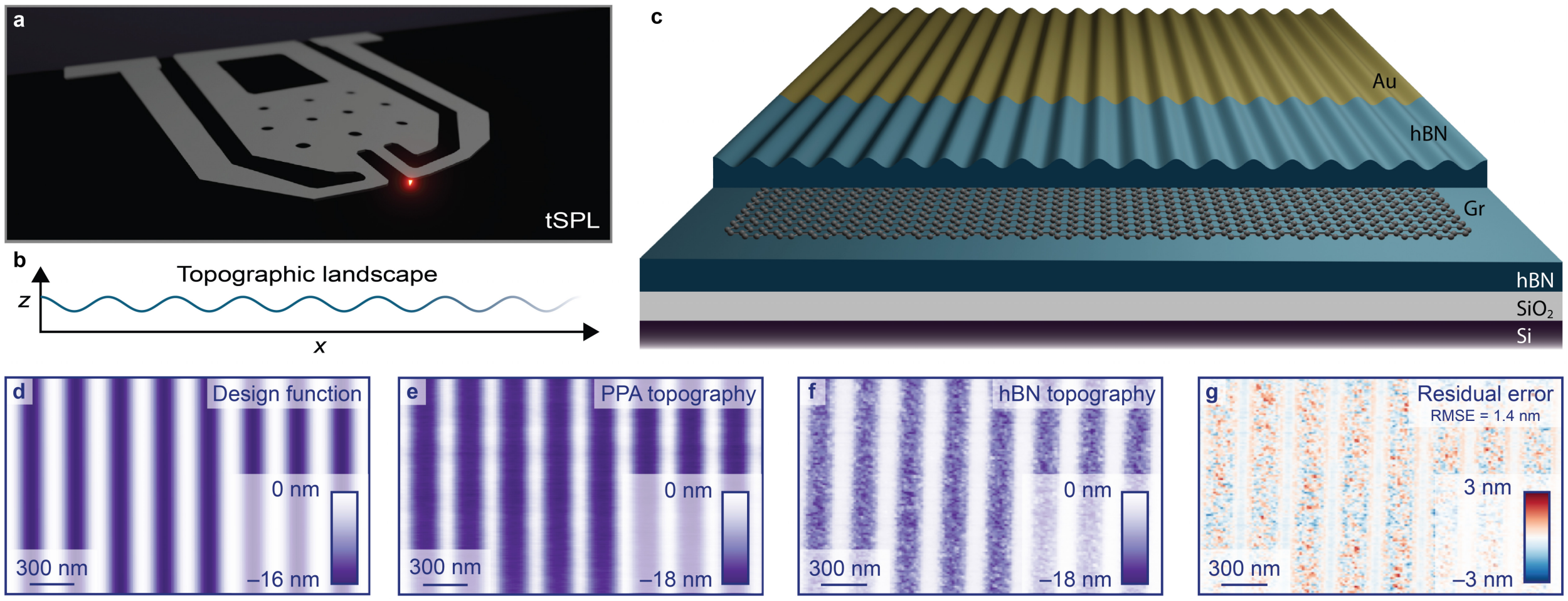

**Figure 1.** Device concept. (a) Rendering of a tSPL cantilever, where the red glow represents a heated tip. (b) The thermal probe patterns a topographic landscape $z(x)$ in a layer of PPA. This landscape is transferred to a hBN flake with RIE. (c) vdW heterostructure showing monolayer graphene encapsulated in hBN, where the top hBN flake contains a patterned landscape, covered with a gold layer. A voltage bias is applied between the graphene and gold layers (top gate) and between the graphene and silicon layers (back gate) to induce charge-carrier doping in graphene. The top hBN flake and gold layer are partially cut in this rendering to clearly show the encapsulated graphene layer. (d) Design function for the sinusoidal landscape mapped on a 20×20 $nm^2$ pixel grid. (e) Measured topography of the patterned landscape in PPA, taken with tSPL during patterning (root-mean-square error: $RMSE_{PPA}$ = 1.3 nm). (f) Topography of the transferred pattern in hBN after RIE, measured with AFM after cleaning. (g) Residual error of the surface profile in hBN compared to the design function, with $RMSE_{hBN}$ = 1.4 nm.

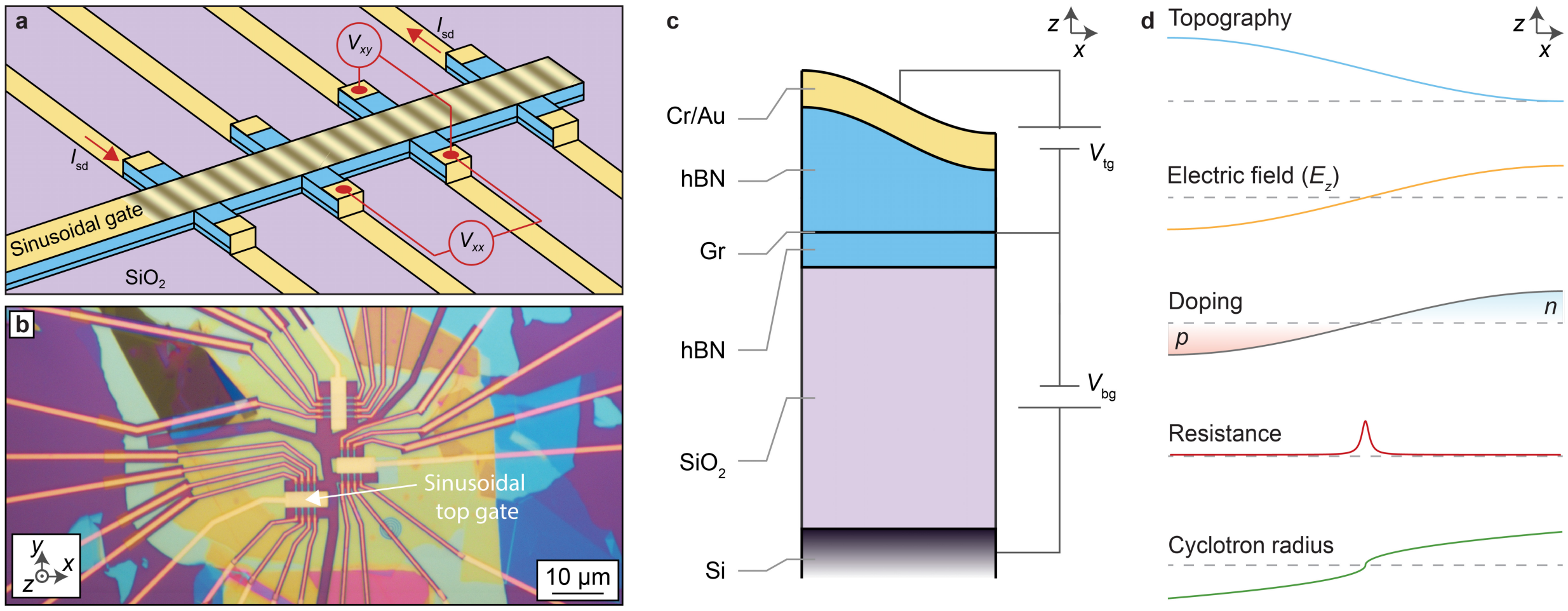

**Figure 2.** Device properties. (a) Schematic of the device, which is a Hall bar with edge contacts and a tSPL-patterned top gate. Current moves from source to drain ($I_{\mathrm{sd}}$) through the device, and the longitudinal ($V_{xx}$) and transverse ($V_{xy}$) voltages can be measured. Doping in the device is controlled by an applied back-gate voltage ($V_{\mathrm{bg}}$) and an applied top-gate voltage ($V_{\mathrm{tg}}$). (b) Optical micrograph of fabricated devices, where the device with a sinusoidal top gate is indicated. (c) Cross-section of the device stack, ranging from the maximum to the minimum depth, illustrating how monolayer graphene is gated with a global back gate and a patterned top gate. (d) Emerging properties vary spatially along the $x$-coordinate when a top-gate voltage is applied. The topographic landscape imposes a gradient electric field at the graphene layer, which affects the spatial distribution of doping, resistance, and cyclotron radius for charges in the device.

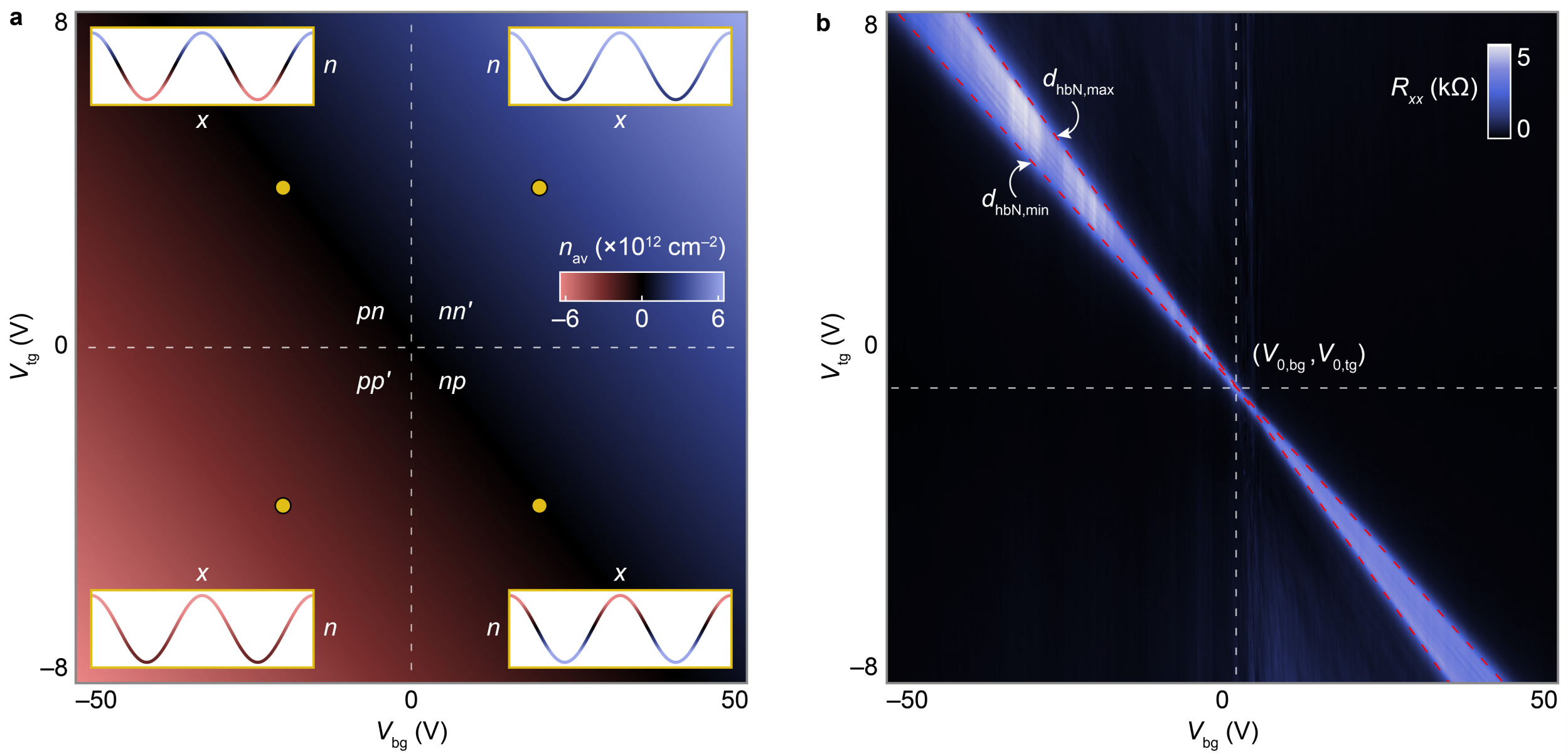

**Figure 3.** Gate-induced doping characteristics. (a) Calculated average carrier density $n_{\text{av}}$ in the device as a function of applied back-gate voltage $V_{\text{bg}}$ and applied top-gate voltage $V_{\text{tg}}$. Four quadrants are marked according to their doping profile, including bipolar *pn* doping with holes and electrons, unipolar *nn'* doping with only electrons, *np* doping with electrons and holes, and *pp'* doping with only holes. The insets show spatially varying carrier density profiles $n(x)$ for a $V_{\text{bg}}$,$V_{\text{tg}}$ pair (yellow dots), normalized to minimum and maximum, where red represents holes and blue represents electrons. (b) Measured $R_{xx}$ as a function of $V_{\text{bg}}$ and $V_{\text{tg}}$ ($T \approx 15$ mK), where the white dashed lines indicate the shifted origin ($V_{0,\text{bg}} = 2.4$ V, $V_{0,\text{tg}} = -1$ V). The red dashed lines indicate the boundaries of resistance-peak spreading, calculated from the minimum and maximum thickness ($d_{\text{hBN,min}}$ and $d_{\text{hBN,max}}$) of the patterned hBN flake that defines the geometry of the top-gate dielectric (Supporting Information).

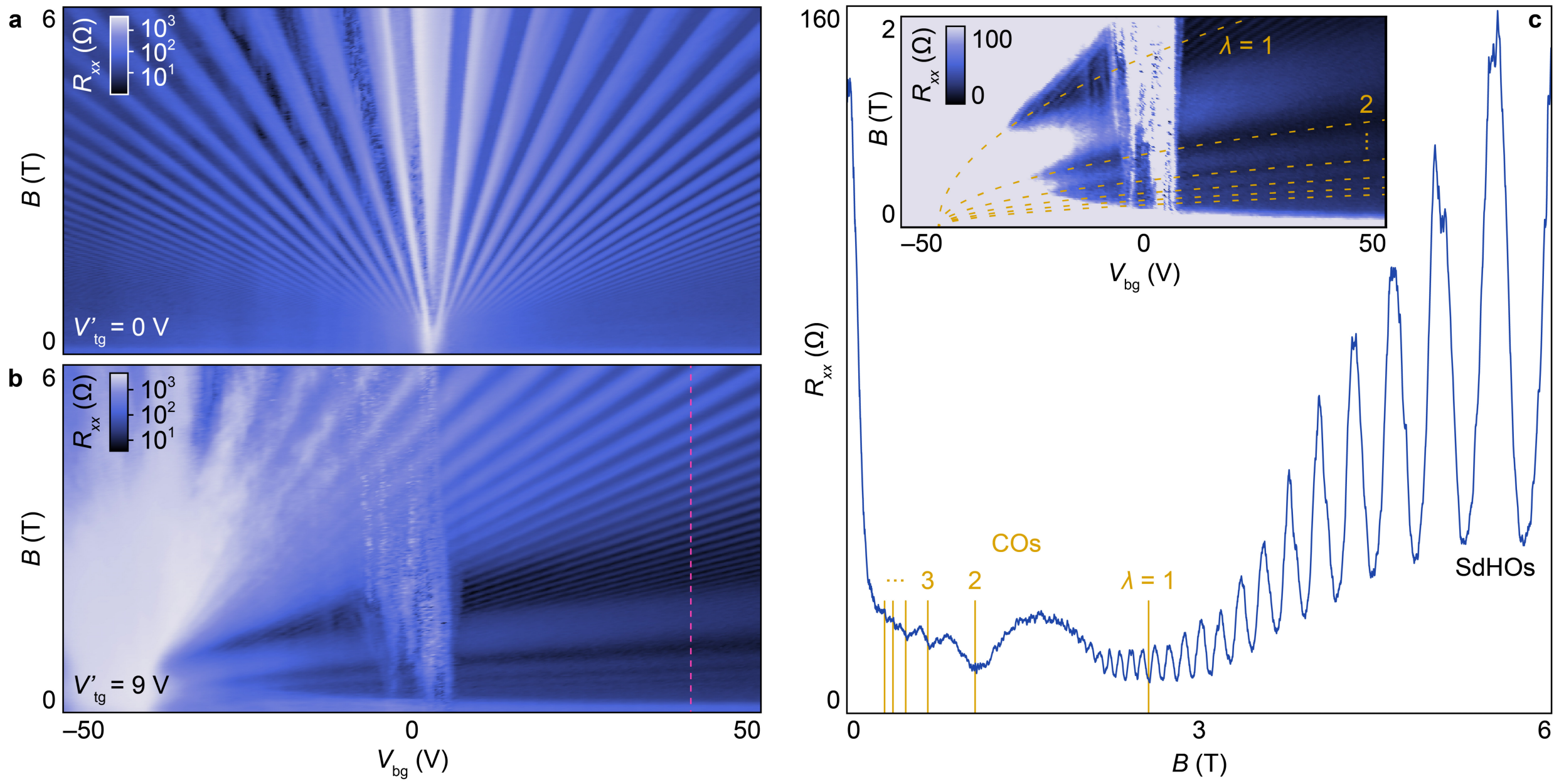

**Figure 4.** Magnetotransport measurements. (a) Measured longitudinal resistance $R_{xx}$ with the top gate set to $V'_{\text{tg}} = 0$ V, as a function of back-gate voltage $V_{\text{bg}}$ and perpendicular magnetic field strength $B$. (b) Measured $R_{xx}$ with the top gate set to $V'_{\text{tg}} = 9$ V, as a function of $V_{\text{bg}}$ and perpendicular magnetic field strength $B$. (c) Line cut of $R_{xx}$ along the $B$-axis at $V'_{\text{tg}} = 9$ V and $V_{\text{bg}} = 45$ V (vertical dashed pink line in panel b), showing commensurability oscillations (COs) and Shubnikov–de Haas oscillations (SdHOs). Yellow lines mark the predicted positions where the commensurability conditions are satisfied, corresponding to dips in $R_{xx}$. The inset shows a zoom-in from panel b, where the dashed yellow lines mark the commensurability conditions given by resistance dips with the expected $B \propto \sqrt{n}$ dependence (Supporting Information). Measurements are performed at $T \approx 15$ mK.

**Table of Contents Graphic**

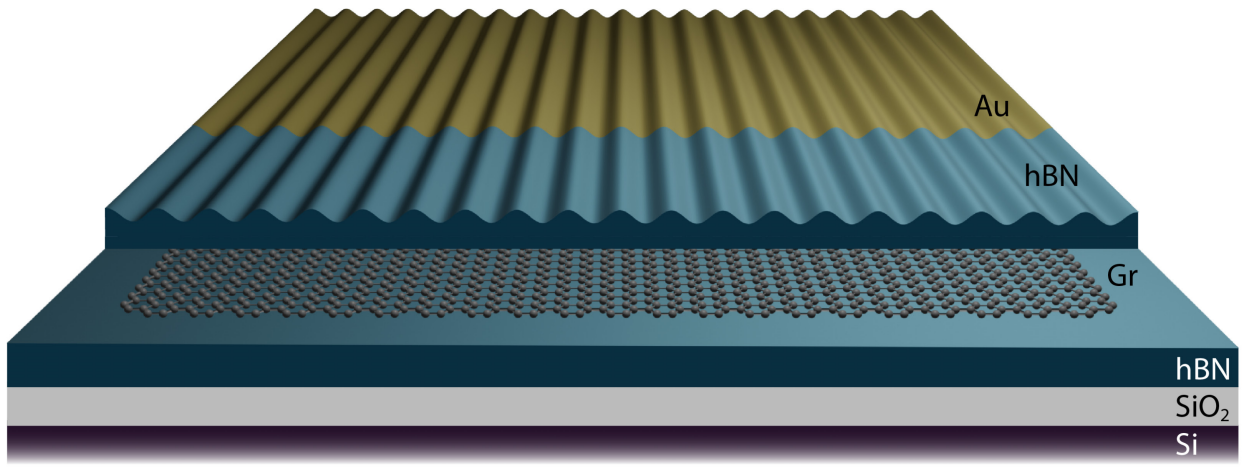

# Supporting Information

## Gradient Electronic Landscapes in van der Waals Heterostructures

*Nolan Lassaline,*[*,†] *Camilla H. Sørensen,*[†] *Giulia Meucci,*[§] *Sander J. Linde,*[†] *Kian Latifi Yaghin,*[†] *Tuan K. Chau,*[†] *Damon J. Carrad,*[§] *Peter Bøggild,*[†] *Thomas S. Jespersen,*[§] *Timothy J. Booth*[†]

[†]Department of Physics, Technical University of Denmark, 2800 Kongens Lyngby, Denmark
[§]Department of Energy Conversion and Storage, Technical University of Denmark, 2800 Kongens Lyngby, Denmark

∗Corresponding author: Nolan Lassaline, E-mail: nlasso@dtu.dk

**Section S1: 2D materials exfoliation and stacking**

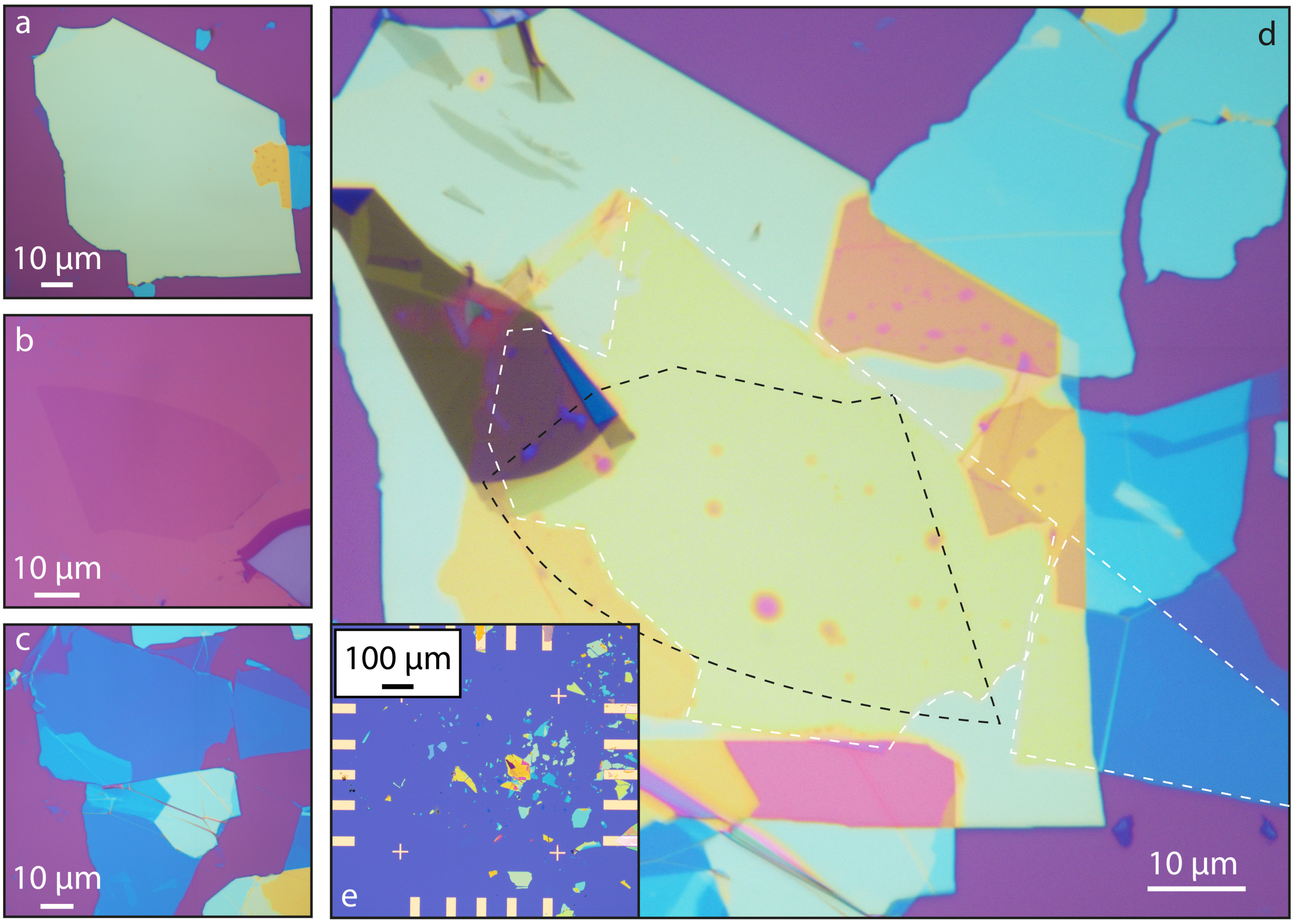

**Figure S1.** Exfoliated crystals and stacked hBN–graphene–hBN heterostructure. (a) Optical microscope image (100×) of the top hBN flake, ~65 nm thick. (b) Optical microscope image (100×) of monolayer graphene. (c) Optical microscope image (100×) of the bottom hBN flake, ~15 nm thick in the uniform region in the top half of the image. (d) Optical microscope image (100×) of the assembled heterostructure, where the location of graphene is indicated with a black dashed line, and the location of the bottom hBN flake is indicated with a white dashed line. (e) Optical microscope image of the chip (10×), showing electrode leads at the edge. This image is modified from reference S1[1].

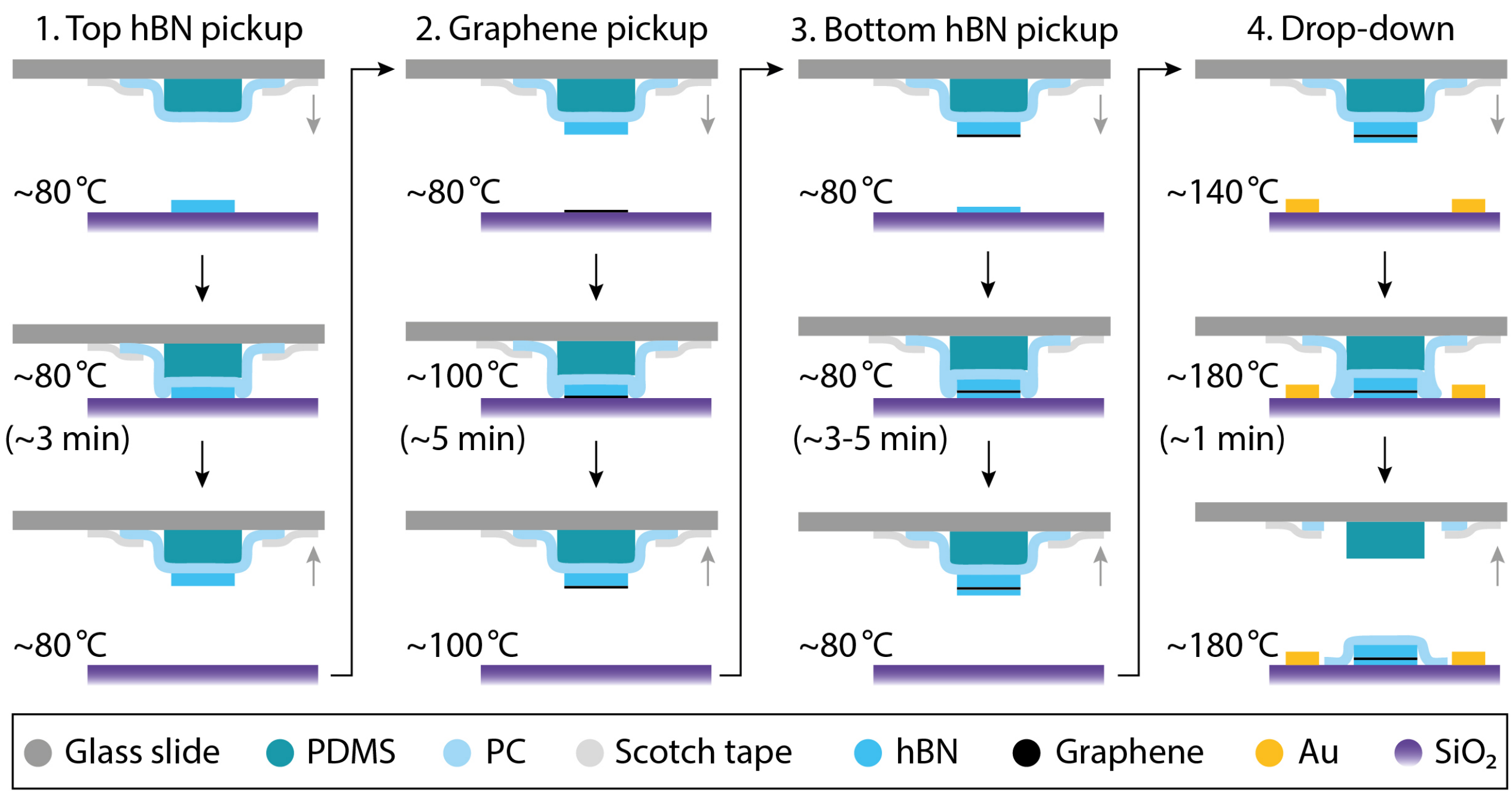

**Figure S2.** Heterostructure stacking. Illustration of the pickup/drop-down steps (from top to bottom) for the top hBN flake (1), monolayer graphene (2), bottom hBN (3), and final drop-down on a Si/$SiO_2$ chip with electrical contact pads (4). After drop-down, the chip is cleaned by dissolving the remaining PC film in chloroform for 2 hours, followed by acetone for 15 minutes, then isopropanol (IPA) for 15 minutes, and finally blown dry with a nitrogen ($N_2$) gun, after which it is ready for lithography. PDMS is polydimethylsiloxane and PC is polycarbonate. This image is modified from reference S1[1].

**Section S2: Device fabrication**

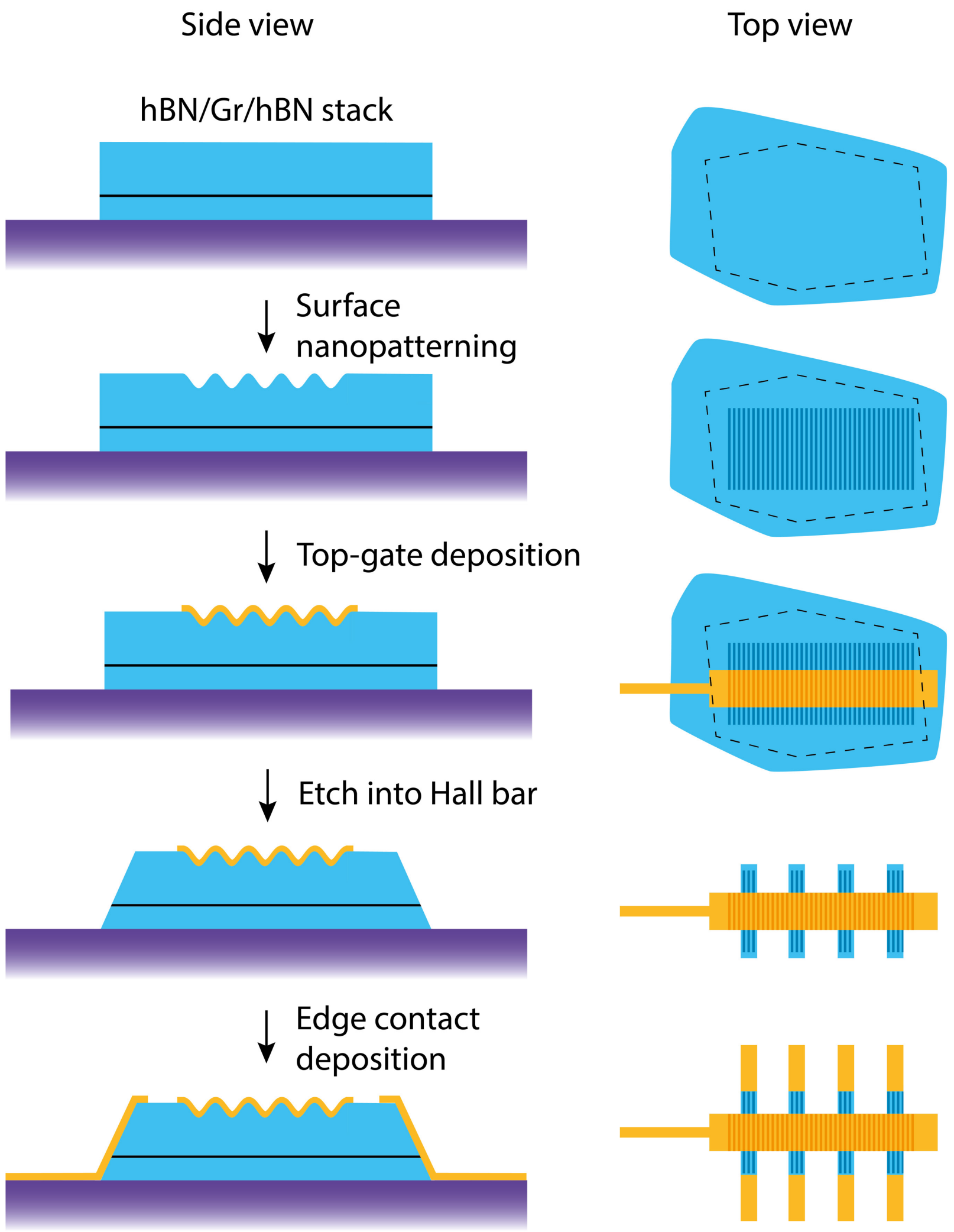

**Figure S3.** Main fabrication steps. Illustration of the overall process flow including the following steps: 1. Heterostructure assembly. 2. Landscape patterning. 3. Top-gate deposition. 4. Device shaping. 5. Contact deposition. The left column is a side view, the right column is the top view. This image is modified from reference S1[1].

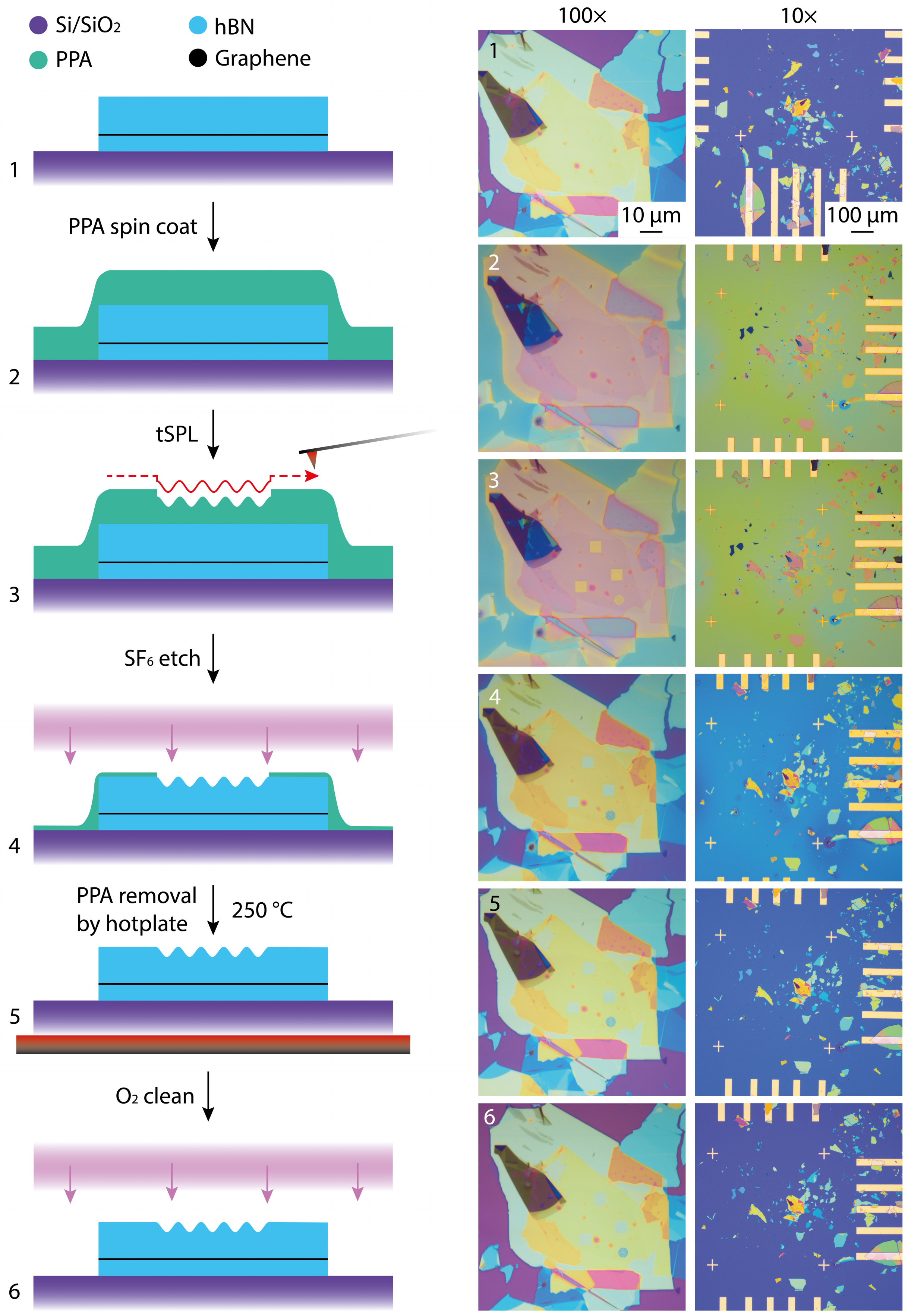

**Figure S4.** Detailed fabrication steps for landscape patterning with tSPL. Illustration (left) and optical microscope images (middle 100×, right 10×) of the landscape patterning and etching process including the following steps: 1. Heterostructure preparation. The sample is treated with $O_2$ plasma (~15 W power, ~10 standard cubic centimeters per minute (sccm) flow rate, ~21 μbar pressure) for 30 s before spin coating to clean the surface and promote resist adhesion. 2. PPA spin coating. A 5 weight per cent (wt %) PPA solution in anisole is spin coated with a speed of 6000 rpm for 40 s, with an acceleration of 500 rpm/s, producing a film of ~ 75 nm thickness. After spin coating, the chip is baked on a hot plate at 110 °C for 2 minutes to remove residual solvent from the PPA layer. 3. tSPL patterning. The temperature setpoint of the probe is adjusted to $T = 1100$ °C, and calibration patterns are performed away from the stack to adjust the feedback controller and set the write forces. The same sine wave (300 nm period, minimum depth 20 nm, maximum depth 36 nm) is used for the calibration and gate pattern to maximize consistency. Once the probe is calibrated, it is used to pattern the region of interest on the stack, resulting in a sine wave with 1.3 nm RMSE. 4. $SF_6$ etching. The sample is etched using $SF_6$ gas (~24 W power, ~15 sccm flow rate, ~6 μbar pressure) for 50 seconds[2]. 5. Excess PPA removal. Excess PPA is removed by placing the chip on a hot plate set to 250 °C for 30 s. 6. $O_2$ plasma cleaning. The chip is cleaned with the same parameters as step 1 above. The final pattern in hBN is measured with AFM, revealing a sine wave with 296 nm wavelength and 12 nm peak-to-valley depth, with a RMSE of 1.4 nm. This image is modified from reference S1[1].

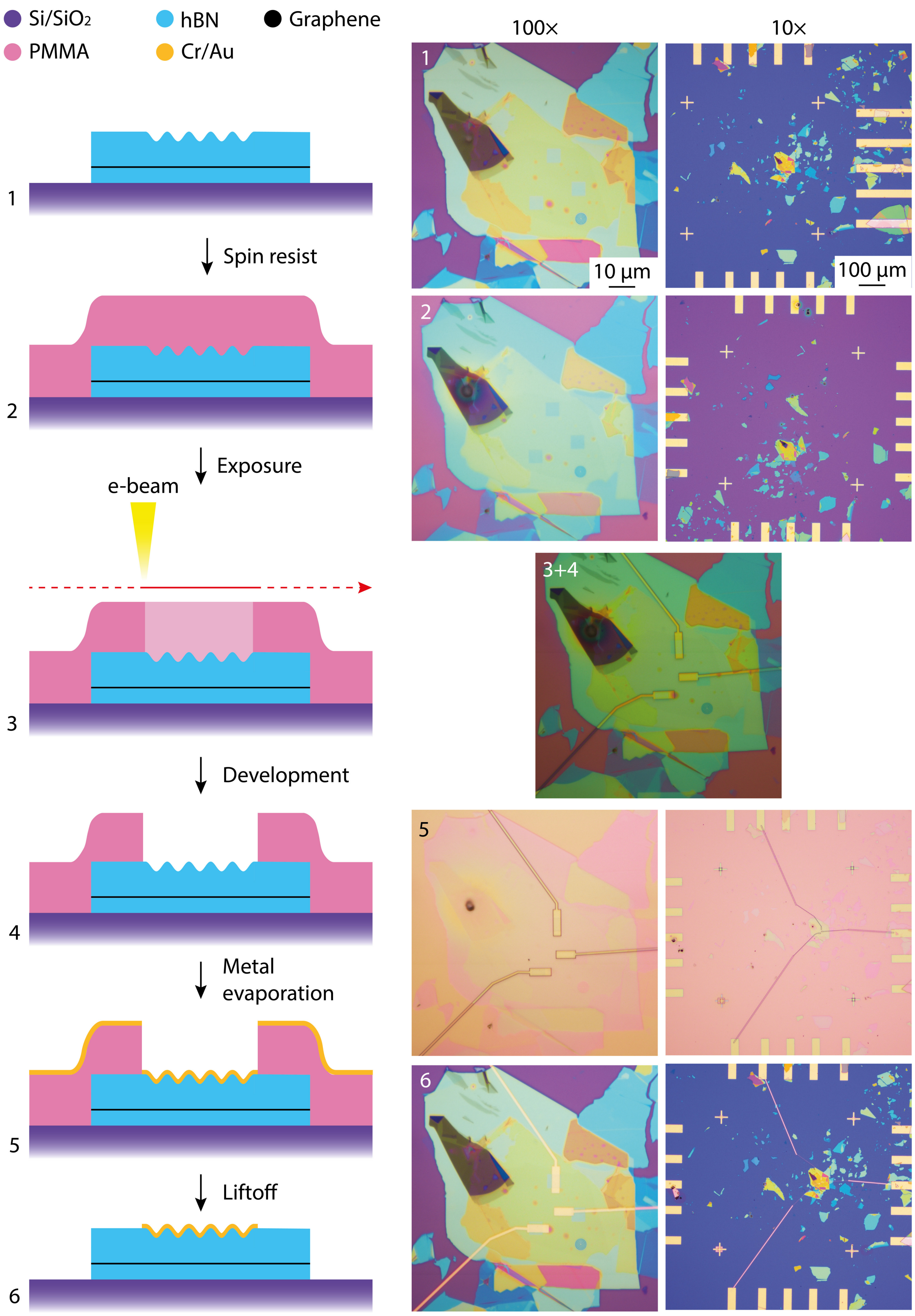

**Figure S5.** Detailed fabrication steps for top-gate metallization. Illustration (left) and optical microscope images (middle 100×, right 10×) of the top-gate patterning and metallization process including the following steps: 1. Heterostructure with patterned surface. 2. PMMA spin coating. A 4 wt % PMMA solution in anisole is spin coated on the chip at 3000 rpm for 60s, with an acceleration of 500 rpm/s, producing a film of ~200 nm thickness. After spin coating, the chip is baked on a hot plate at 150 °C for 2 minutes to remove residual solvent from the PMMA layer. 3. Electron beam lithography (EBL). The chip is exposed with a beam current of 1 nA and a dose of 1000 $\mu C/cm^2$, using optical microscope images to design and align the exposure pattern. 4. Development. The resist is developed by immersing the chip in a 1:3 MIBK:IPA solution for 20 s to dissolve the exposed PMMA regions (MIBK is methyl isobutyl ketone). Afterwards, the chip is directly immersed in pure IPA for 20 s. Finally, the chip is blown dry with a $N_2$ gun. 5. Cr/Au thermal evaporation of ~5 nm/~50 nm. Cr was evaporated from a rod using a set current of ~92 A, resulting in a rate of ~0.1 Å/s. Au was evaporated from a crucible using a set current of ~41 A, resulting in a deposition rate of ~0.5 Å/s. 6. Liftoff process. Liftoff is performed by placing the chip in acetone on a hotplate set to 60 °C for ~2 hours, after which warm acetone is pipetted over the surface of the chip to remove loose gold. The chip is then rinsed in IPA and blown dry with a $N_2$ gun. This image is modified from reference S1[1].

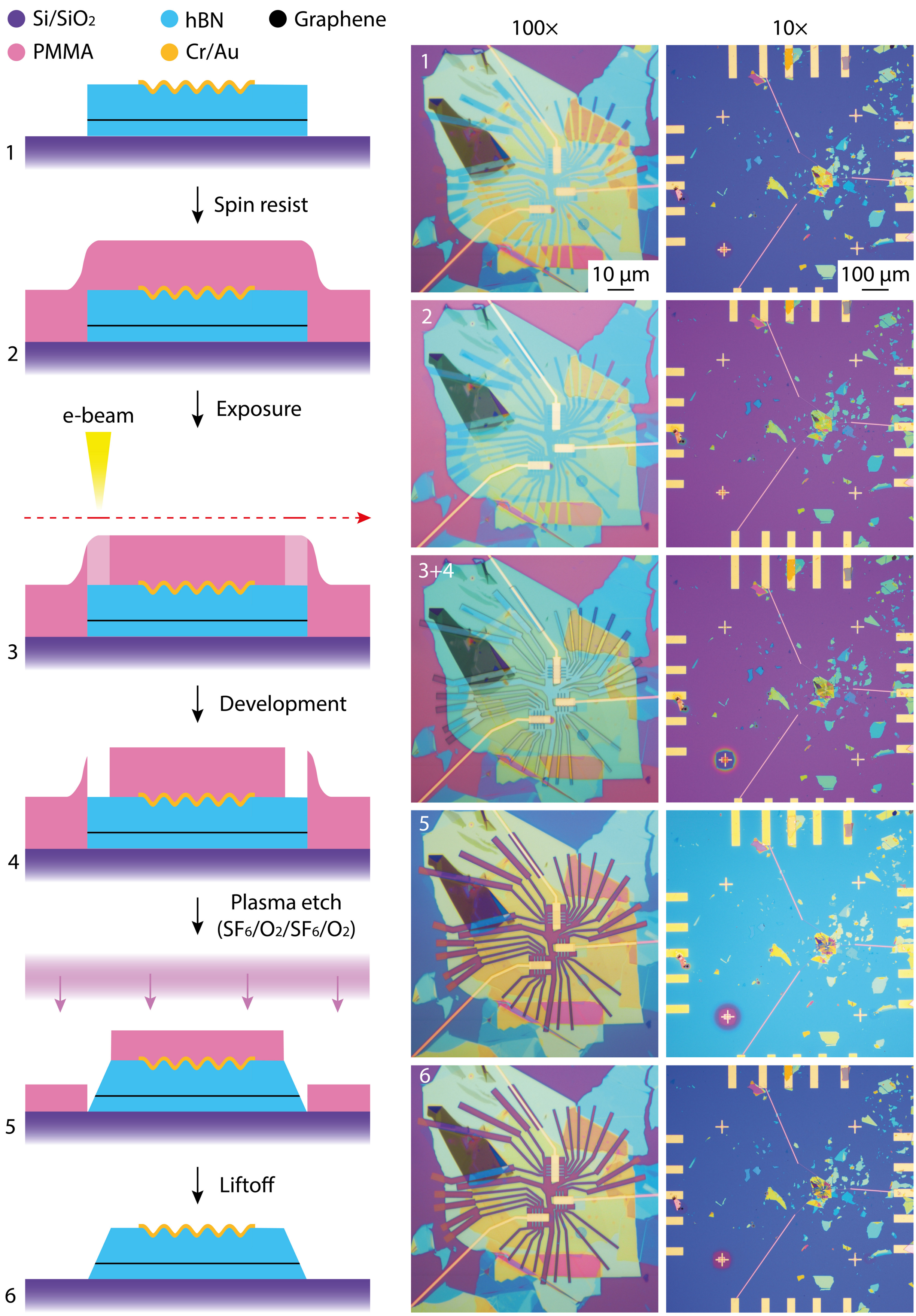

**Figure S6.** Detailed fabrication steps for device shaping. Illustration (left) and optical microscope images (middle 100×, right 10×) of the patterned Hall bar and electrical isolation regions, and the etching process, including the following steps: 1. Heterostructure with local metal top gates. 2. PMMA spin coating. 3. EBL patterning. 4. Development. Process parameters and chemical reagents in steps 2–4 are the same as described in Figure S5. 5. Etching the hBN/graphene/hBN stack with alternating $SF_6$/$O_2$/$SF_6$/$O_2$ plasma steps with times of 46 s/29 s/11 s/28 s, using the same etch parameters for $SF_6$ gas and $O_2$ gas as Figure S4. 6. Chip cleaning (same as Figure S5). This image is modified from reference S1[1]. We note that our initial attempt at these steps failed with an incomplete etch due to an unexpectedly lower rate for that particular attempt, such that we had to spin coat resist and perform lithography and etching steps again to bring the etch to completion all the way through the stack, resulting in a partially etched pattern that is visible in the top row.

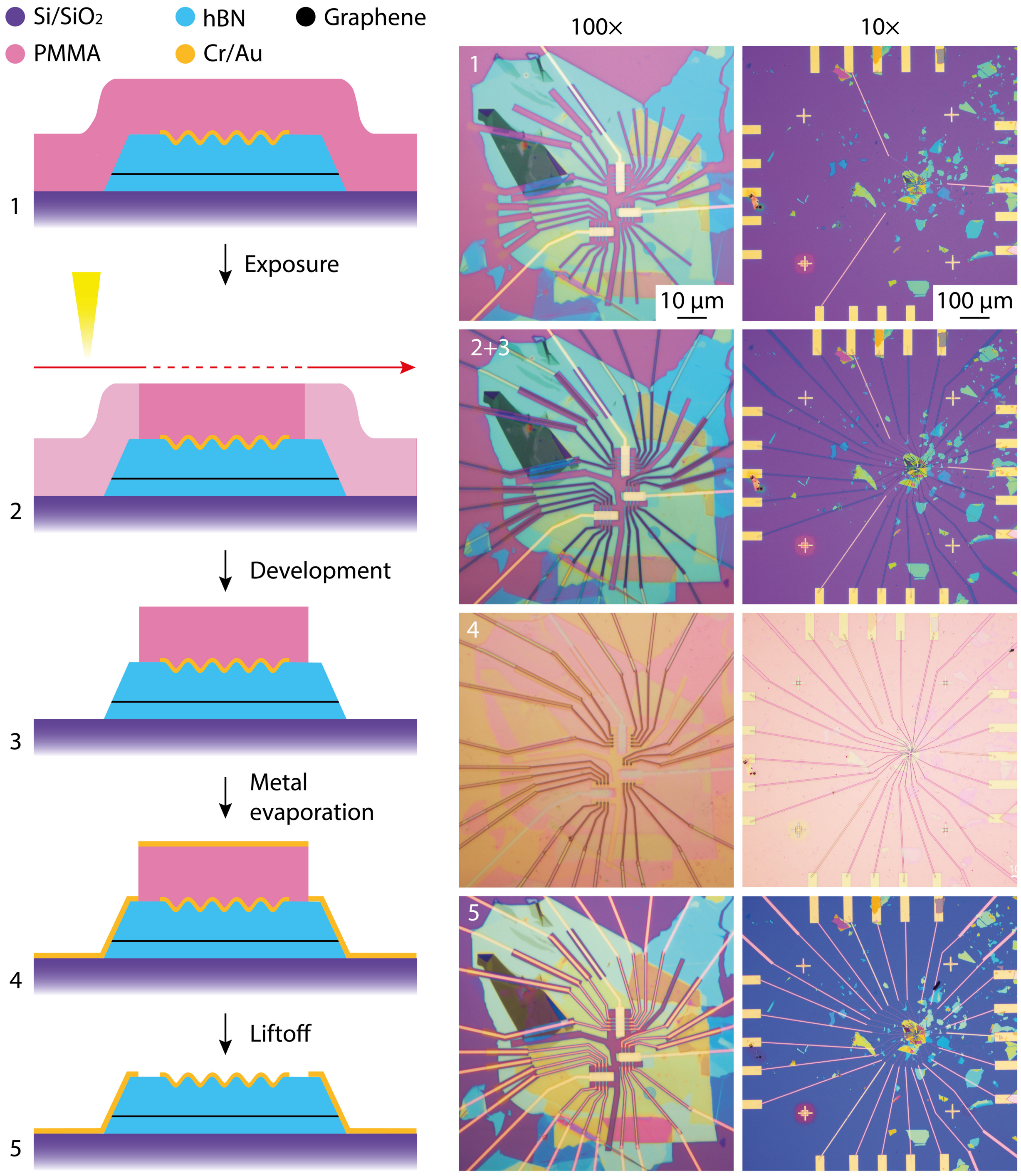

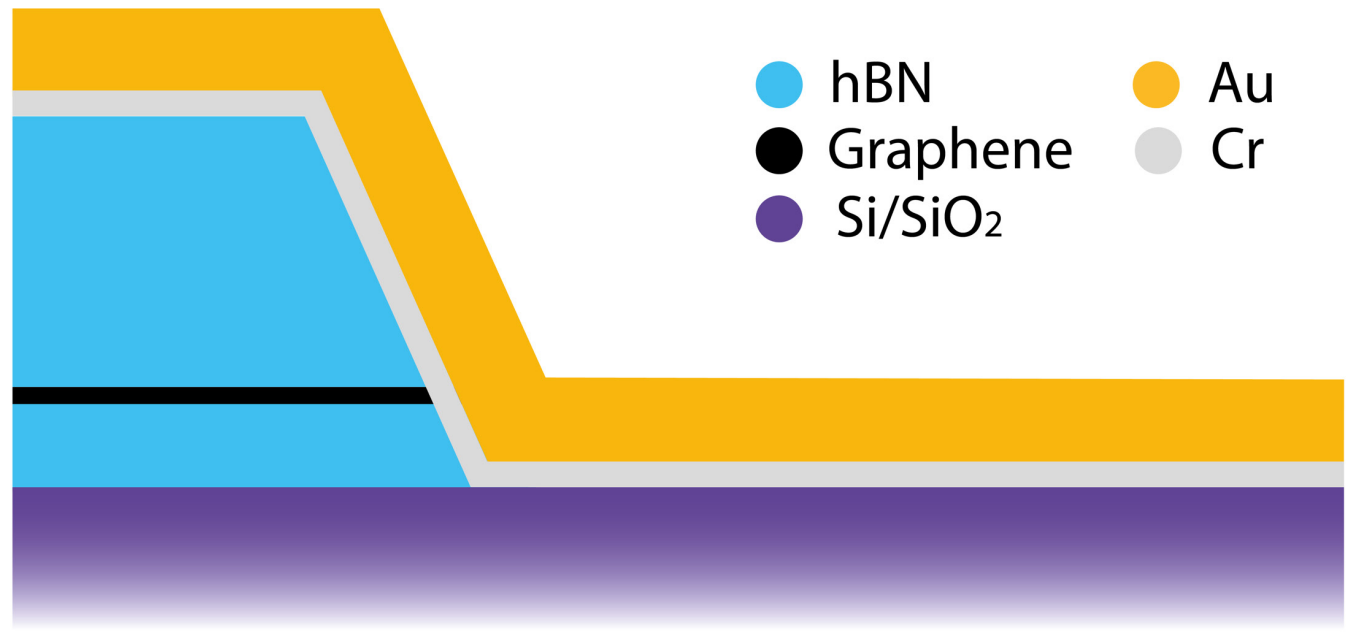
hBN
Au
Graphene
Cr
Si/SiO2

**Figure S7.** Detailed fabrication steps for edge contacting. Illustration (left) and optical microscope images (middle 100×, right 10×) of the contact patterning and metallization, including the following steps: 1. PMMA spin coating. 2. EBL patterning. 3. Development. 4. Cr/Au thermal evaporation. 5. Liftoff process. 6. Chip cleaning. Steps 1–5 are the same as described in Figure S5. The bottom panel is a cross-section illustrating the geometry of edge contacts. This image is modified from reference S1[1].

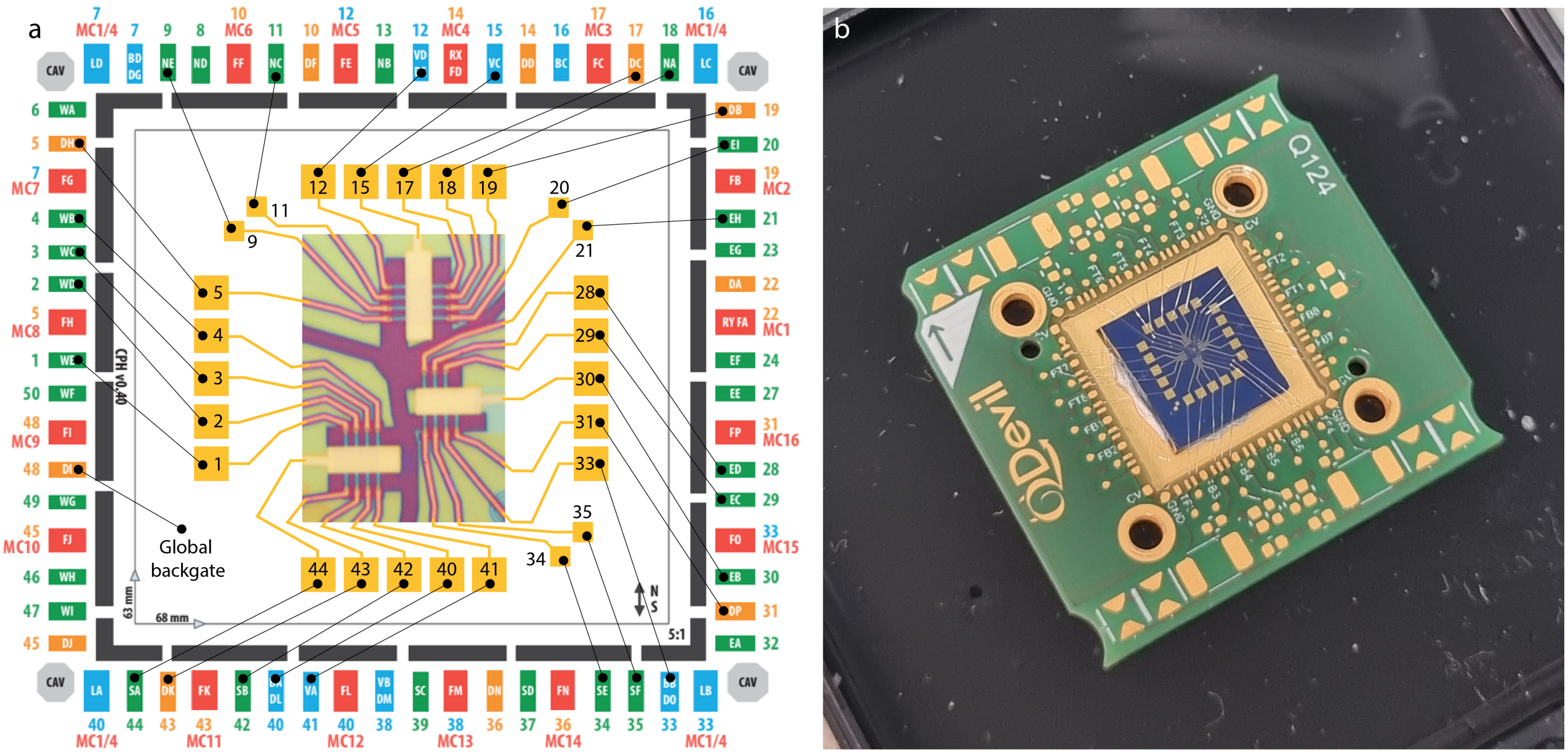

**Figure S8.** Chip layout. (a) Overview of the wire-bonded chip. Yellow lines and squares indicate electrodes on the chip, and black lines indicate wire-bonded electrical connections from contact pads on the chip to electrodes on the carrier motherboard (green, orange, blue rectangles). (b) Photograph of the chip (blue) wire-bonded and fixed to the carrier (green/gold) with conductive silver paste. This allows electrical access to the silicon plane of the chip to act as a global back-gate electrode. This image is modified from reference S1[1].

**Section S3: Modelling gate-induced charge carrier doping in graphene**

Each gate–graphene combination acts as a parallel plate capacitor. For a capacitor with plate area $A$, dielectric thickness $d$, and relative permittivity $\varepsilon_\mathrm{r}$, the classical capacitance is given by:

$$C = \frac{\varepsilon_0 \varepsilon_\mathrm{r} A}{d} \tag{S1}$$

Where $\varepsilon_0$ is the vacuum permittivity. When a voltage bias $V$ is applied as a potential difference between the two plates, the surface charge density on each plate is given by:

$$\sigma_\mathrm{s} = C\frac{V}{A} = \frac{\varepsilon_0 \varepsilon_\mathrm{r}}{d}V = C_\mathrm{g} V \tag{S2}$$

where $C_\mathrm{g}$ is the capacitance per unit area for the gate. The device in this work has both a back gate and a top gate ($C_\mathrm{bg}$ for the back gate and $C_\mathrm{tg}$ for the top gate), which each have a contribution to the surface charge density, given by:

$$\sigma_\mathrm{bg} = \frac{\varepsilon_0 \varepsilon_\mathrm{r,bg}}{d_\mathrm{bg}}\left(V_\mathrm{bg} - V_\mathrm{bg,0}\right) = C_\mathrm{bg} V'_\mathrm{bg} \tag{S3}$$

$$\sigma_\mathrm{tg} = \frac{\varepsilon_0 \varepsilon_\mathrm{r,tg}}{d_\mathrm{tg}}\left(V_\mathrm{tg} - V_\mathrm{tg,0}\right) = C_\mathrm{tg} V'_\mathrm{tg} \tag{S4}$$

where $V_\mathrm{bg}$ is the applied back-gate voltage (between graphene and silicon) and $V_\mathrm{tg}$ is the applied top-gate voltage (between graphene and gold). The applied voltages are shifted by $V_\mathrm{bg,0}$ and $V_\mathrm{tg,0}$ possibly due to charge-impurity doping, where we define the shifted axes as $V'_\mathrm{bg} = V_\mathrm{bg} - V_\mathrm{bg,0}$ and $V'_\mathrm{tg} = V_\mathrm{tg} - V_\mathrm{tg,0}$. The carrier density $n$ is given by dividing the total charge density $\sigma_\mathrm{tot} = \sigma_\mathrm{bg} + \sigma_\mathrm{tg}$ by the elementary charge $e$:

$$n = \frac{\sigma_{\text{tot}}}{e} = \frac{1}{e}\left(C_{\text{bg}}V'_{\text{bg}} + C_{\text{tg}}V'_{\text{tg}}\right) \tag{S5}$$

The device in this work has a top gate with a sinusoidally varying thickness along the $x$-axis, given by:

$$d_{\text{tg}}(x) = d_{\text{av}} + d_{\text{mod}}\sin\left(\frac{2\pi}{a}x\right) \tag{S6}$$

where $d_{\text{av}}$ is the average thickness, $d_{\text{mod}}$ is the amplitude describing the thickness modulation, and $a$ is the periodicity of the modulation. The result of this topographic modulation is a carrier density spatially modulated along the $x$-axis, given by:

$$n(x) = n_{\text{bg}} + n_{\text{tg}}(x) = n_{\text{bg}} + \frac{\varepsilon_0\varepsilon_{\text{r,tg}}}{e\left[d_{\text{av}} + d_{\text{mod}}\sin\left(\frac{2\pi}{a}x\right)\right]}V'_{\text{tg}} \tag{S7}$$

with

$$n_{\text{bg}} = \frac{\varepsilon_0\varepsilon_{\text{r,bg}}}{ed_{\text{bg}}}V'_{\text{bg}} = C_{\text{bg}}V'_{\text{bg}} \tag{S8}$$

where $n_{\text{bg}}$ is the contribution from the back gate to the carrier density, and $n_{\text{tg}}(x)$ is the spatially modulated contribution from the patterned top gate to the carrier density. To model the resistance-peak spreading in Figure 3b, lines marking the boundaries for charge neutrality can be calculated from the minimum and maximum hBN thickness. The two thickness extrema produce two charge neutrality conditions where the induced carrier densities from the top gate and back gate are equal in magnitude and opposite in sign $n_{\text{bg}} = -n_{\text{tg}}$, given by:

$$V'_{\mathrm{tg}} = -\frac{d_{\mathrm{tg,min}}}{\varepsilon_{\mathrm{r,tg}}}\frac{\varepsilon_{\mathrm{r,bg}}}{d_{\mathrm{bg}}}V'_{\mathrm{bg}} \tag{S9}$$

$$V'_{\mathrm{tg}} = -\frac{d_{\mathrm{tg,max}}}{\varepsilon_{\mathrm{r,tg}}}\frac{\varepsilon_{\mathrm{r,bg}}}{d_{\mathrm{bg}}}V'_{\mathrm{bg}} \tag{S10}$$

These two relationships are plotted as red dashed lines in Figure 3b, marking boundaries between which the device has high resistance due to the peak existing inside the device for a range of $V'_{\mathrm{bg}}$ that depends on $V'_{\mathrm{tg}}$, which we call resistance-peak spreading. The calculation uses the quantities: $\varepsilon_{\mathrm{r,bg}} = \varepsilon_{\mathrm{r,SiO_2}} \approx 3.9$, $\varepsilon_{\mathrm{r,tg}} = \varepsilon_{\mathrm{r,hBN\perp}} \approx 3.8$, $d_{\mathrm{bg}} = 300\ \mathrm{nm}$, $d_{\mathrm{tg,max}} = d_{\mathrm{hBN,max}} = 65\ \mathrm{nm}$, and $d_{\mathrm{tg,min}} = d_{\mathrm{hBN,min}} = 65\ \mathrm{nm} - 2d_{\mathrm{mod}} - \mathrm{RMSE} = 65\ \mathrm{nm} - 12\ \mathrm{nm} - 1\ \mathrm{nm} = 52\ \mathrm{nm}$. Here we subtract the RMSE (rounded to 1 nm) to include deviations from an ideal sine wave due to imperfections, where the calculated $d_{\mathrm{tg,min}}$ captures these differences with a slight reduction in thickness due to roughness. We note that while the back gate is composed of the series combination of ~285 nm $SiO_2$ and 15 nm hBN, the similar values for permittivity $\varepsilon_{\mathrm{r,SiO_2}} \approx \varepsilon_{\mathrm{r,hBN\perp}}$ allow us to approximate the back gate as 300 nm of $SiO_2$ for simplicity.

## Section S4: Determining mobility

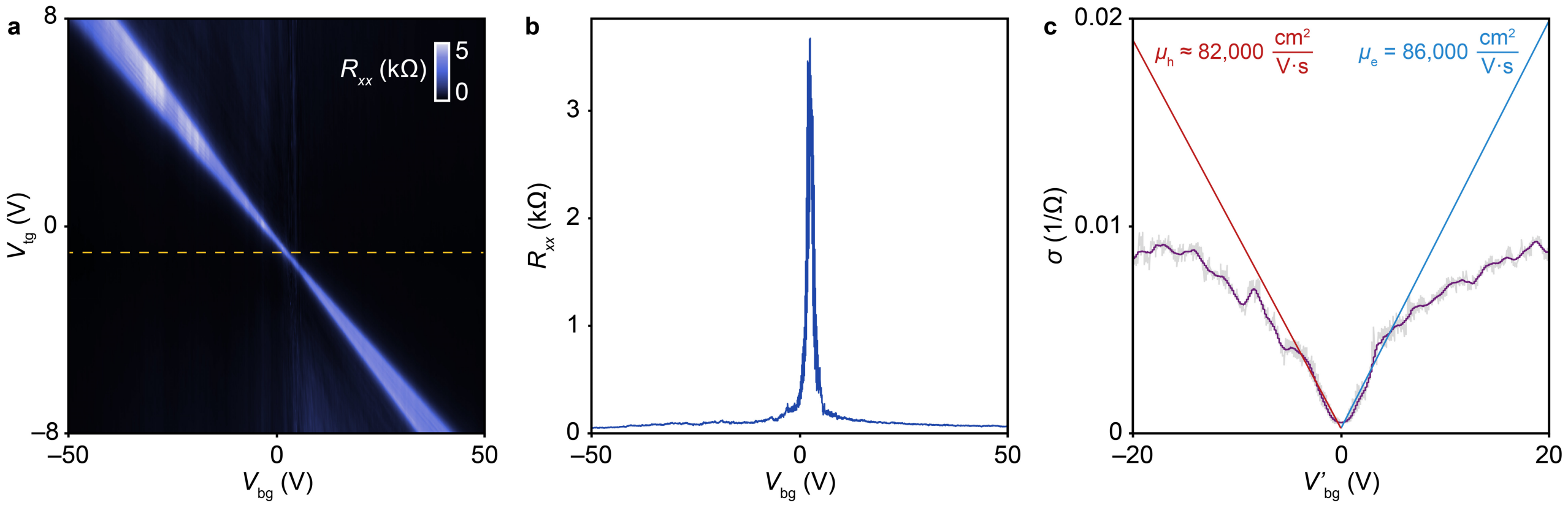

**Figure S9.** Determining the field-effect mobility. (a) Measured $R_{xx}$ as a function of $V_{\mathrm{bg}}$ and $V_{\mathrm{tg}}$, where the dashed yellow line indicates $V'_{\mathrm{tg}} = 0$. (b) Line cut along the yellow dashed line in panel a, showing the resistance peak without direct influence from the top gate. (c) Conductivity $\sigma$ calculated from the data in panel b, plotted as a function of shifted back-gate voltage $V'_{\mathrm{bg}}$. The slope of the blue (red) line can be used to determine the field-effect mobility for electrons (holes)[3], which is obtained by fitting a linear function to the linear region of a smoothed version (purple, moving average with 1.09 V width) of the raw data (gray), revealing $\mu_{\mathrm{e}} \approx 86{,}000 \frac{\mathrm{cm}^2}{\mathrm{V \cdot s}}$ and $\mu_{\mathrm{h}} \approx 82{,}000 \frac{\mathrm{cm}^2}{\mathrm{V \cdot s}}$.

**Section S5: Modelling commensurability dips**

The expected position for dips in $R_{xx}$ corresponding to COs can be predicted if the carrier density $n$ is known. We exploit the fact that SdHOs are periodic in reciprocal magnetic field $1/B$, allowing us to extract the carrier density by determining the oscillation wavelength $\Lambda_{\text{SdHO}}$. We first plot the measured $R_{xx}$ as a function of $V_{\text{bg}}$ and $1/B$, shown in Figure S10a. The oscillations are periodic along the $1/B$ axis with a varying period as a function of $V_{\text{bg}}$. We then take the Fast Fourier Transform (FFT) of the data in Figure S10a along the $1/B$-axis, allowing us to isolate the peak corresponding to the SdHOs and determine the wavelength $\Lambda_{\text{SdHO}}$ (Figure S10b). Carrier density is related to the wavelength by the following expression:

$$n = \frac{2e}{\pi \hbar \Lambda_{\text{SdHO}}} \tag{S11}$$

This allows us to extract carrier density from SdHOs for all relevant $V_{\text{bg}}$. For example, under the conditions in Figure 4c ($V_{\text{bg}} = 45$ V, $V'_{\text{tg}} = 9$ V), we extract $n = 5.8 \times 10^{12}$ cm$^{-2}$, consistent with our theoretical calculations. We can then predict the positions of CO dips on the $R_{xx}(V_{\text{bg}}, B)$ map, based on the following expression:

$$B = \frac{\hbar}{e} \frac{2\sqrt{\pi}}{\left(\lambda - \frac{1}{4}\right) a} \sqrt{n} \tag{S12}$$

Where $\lambda$ is an integer and $a$ is the periodicity of the gated potential. The results are plotted in Figure 4c (yellow lines), showing an excellent agreement with the position of the measured dips.

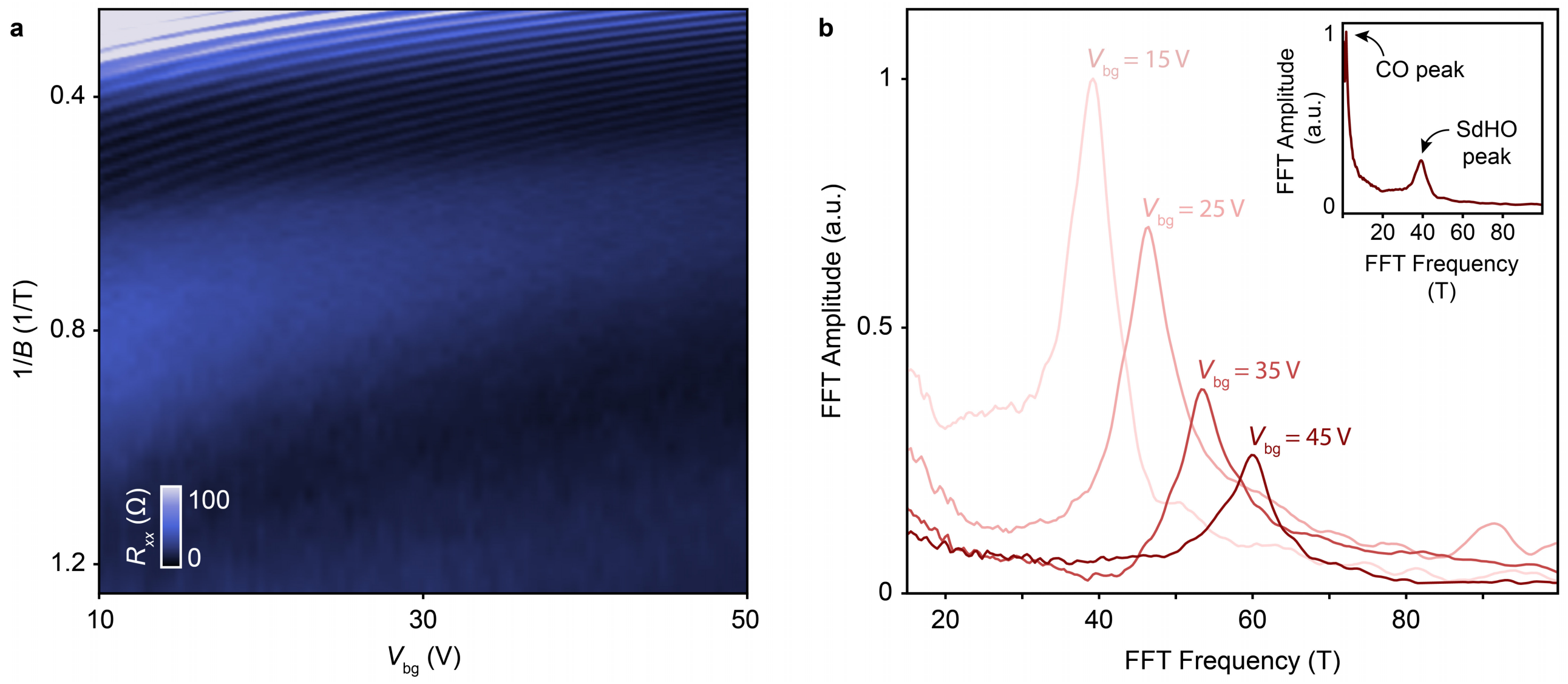

**Figure S10.** Oscillation analysis. (a) Longitudinal resistance $R_{xx}$ plotted as a function of back-gate voltage $V_{\mathrm{bg}}$ and reciprocal magnetic field $1/B$ for ~2 CO periods, where COs can be seen as low-frequency oscillations at high $1/B$ values, and SdHOs can be seen as high-frequency oscillations at low $1/B$ values. (b) Line cuts from the Fast Fourier Transform (FFT) of the data in panel a along the $1/B$ axis, zoomed in to the peak corresponding to SdHOs plotted for a series of constant $V_{\mathrm{bg}}$ (red lines). These peak positions provide information on the oscillation wavelength $\Lambda_{\mathrm{SdHO}}$, which are determined for all relevant $V_{\mathrm{bg}}$ to extract carrier density with Equation S11, allowing us to calculate the dashed yellow curves in Figure 4c with Equation S12. The inset shows the entire range of the FFT, where COs contribute to the low-frequency peak and SdHOs contribute to the high-frequency peak.

**Supplementary References**